\documentclass[12pt]{article}
\usepackage{amsmath}
\usepackage{times}
\usepackage{graphicx}
\usepackage{color}
\usepackage{multirow}
\usepackage[authoryear]{natbib}
\usepackage{rotating}
\usepackage{bbm}
\usepackage{latexsym}
\usepackage[american,siunitx]{circuitikz}
\usetikzlibrary{shapes.geometric, decorations.pathreplacing,calc,positioning,arrows.meta}
\tikzstyle{LTI} = [rectangle, minimum width=2cm, minimum height=1cm,text centered, draw=black]
\tikzstyle{a} = [rectangle, minimum width=1cm, minimum height=1cm,text centered, draw=black]
\tikzstyle{arrow} = [thick,->,>=stealth, line cap = round]
\tikzstyle{box} = [rectangle, minimum width=1cm, minimum height=1cm,text centered, draw=black]
\tikzstyle{block} = [rectangle, minimum width = 1cm, minimum height = 1cm, text centered, draw=black]
\tikzstyle{round} = [node distance=1.0cm, circle, minimum size=4.5mm, draw, inner sep=0pt]

\renewcommand{\thesubsubsection}{\thesubsection.\arabic{subsubsection}}

\newcommand{\captionfonts}{\normalsize}

\makeatletter  
\long\def\@makecaption#1#2{%
  \vskip\abovecaptionskip
  \sbox\@tempboxa{{\captionfonts #1: #2}}%
  \ifdim \wd\@tempboxa >\hsize
    {\captionfonts #1: #2\par}
  \else
    \hbox to\hsize{\hfil\box\@tempboxa\hfil}%
  \fi
  \vskip\belowcaptionskip}
\makeatother   


\begin{document}

\ \vspace{20mm}\\

{\LARGE Excitability across scales: a unifying EI model}

\ \\
{\bf \large Thomas S.\,J. Burger$^{\displaystyle 1}$,  Rodolphe Sepulchre$^{\displaystyle 1,2}$}\\
{$^{\displaystyle 1}$ STADIUS, Department of Electrical Engineering, KU Leuven, Leuven, Belgium. \texttt{thomas.burger@kuleuven.be, rodolphe.sepulchre@kuleuven.be}\\
{$^{\displaystyle 2}$Department of Engineering, University of Cambridge,
Trumpington Street, Cambridge CB2 1PZ  \texttt{rs771@cam.ac.uk}}}\\
%


\thispagestyle{empty}
\ \vspace{-0mm}\\
%
\begin{center} {\bf Abstract} \end{center}
Neuronal excitability is commonly modeled using different principles at the cellular and population scales, making it difficult to study how microscopic mechanisms shape macroscopic activity. We introduce a spatiotemporal excitatory–inhibitory (EI) model that unifies these scales through a common mixed-feedback architecture. The model retains the conductance-based circuit structure of Hodgkin–Huxley models and describes intrinsic and synaptic currents through memductances whose temporal and spatial voltage dependence encodes dynamics at the relevant scale. In its minimal form, temporal excitability arises from fast excitatory feedback opposed by slower inhibitory feedback, while spatial excitability arises from short-range excitation opposed by longer-range inhibition, recovering the core lateral-inhibition motif of Amari neural fields.
We show that combining these mechanisms gives rise to population excitability in strongly heterogeneous populations. In particular, the model generates population spikes that are localized in both space and time and possess population-level scales distinct from those of individual neuronal spikes, despite substantial cell-to-cell parameter variability. We then show that both intrinsic temporal excitability and spatial synaptic feedback are required for coherent mean-field events and robust spatiotemporal population rhythms. Finally, selective modulation of memductances in a subset of the population tunes the frequency of the population rhythm, illustrating how changes at the cellular scale can shape macroscopic dynamics.
By separating circuit architecture from modeling scale, the proposed framework provides a physically interpretable approach to analyzing and designing neuronal excitability across scales.

\section{Introduction}
Neuronal circuits function on spatial and temporal scales that span several orders of magnitude,
from milliseconds to years, and from micrometers to dozens of centimeters.
As just one example, neuronal activity in the brain displays synchrony across multiple temporal and spatial scales,
which has implications for a range of functions~\citep{kahanaCognitiveCorrelatesHuman2006}.

Modeling excitability has been a key focus of mathematical modeling in neuroscience since the seminal work of \citet{hodgkinQuantitativeDescriptionMembrane1952} and the  recent textbooks \citep{keener2009mathematical,izhikevich2007dynamical,ermentrout2010foundations}. 
The tradition has been, however, to adopt different modeling principles to capture excitability at different scales. 
The  biophysical principles of Hodgkin and Huxley have been broadly adopted to model excitability at the level of a single neuron or for small circuits involving just a few neurons \citep[e.g.,][]{buchholtz1992mathematical}. At the same time, modeling excitability at the level of neuronal populations has been addressed with mean-field approaches (e.g.\ the model of  \citet{wilsonExcitatoryInhibitoryInteractions1972} and the model of \citet{amariDynamicsPatternFormation1977}). While cellular and population models of excitability share a number of qualitative properties, they are nevertheless of a different nature. Both cellular and mean-field models have been successful at describing neuronal behaviors at distinct scales, but the separate nature of the models has made it difficult to study excitability {\it across} scales, that is, to explore how  cellular excitability mechanisms shape and modulate population excitability mechanisms. Attempts have been made to 
derive the macroscopic behavior  from single-cell properties \citep[see, e.g.,][]{coombesNextGenerationNeural2019},
but at the price of starting from abstract neuron models that are difficult to relate to biophysical modeling principles.

Two previous works by the authors, in particular \citet{drion2018switchable} and the recent PhD dissertation by \citet{burger_thesis}, illustrate the shortcomings of modeling cellular and population activity with different modeling principles. 
\citet{drion2018switchable} show how neuromodulators can shape spatiotemporal activity (``brain states'') by modulating the intrinsic properties of the neurons rather than the synaptic neural connections. \citet{burger_thesis} demonstrates how particular properties at the cellular scale determine the robustness of the mean-field PING rhythm at the population scale. In both examples, a population phenomenon is regulated at the cellular scale, and the only way to study the mechanism ``across scales'' is to simulate the entire population at the neuronal scale, which is cumbersome and not scalable. 

Motivated by those shortcomings, we aim to develop new population models that will make it easier to study neuronal behaviors across scales. As a first step in that direction, we propose a spatiotemporal model that unifies the temporal model of Hodgkin and Huxley and the spatial model of Amari. 

Our angle of attack is to retain the key biophysical modeling principles of Hodgkin and Huxley, namely the circuit representation of each neuron as the parallel connection of a capacitive element with a bank of voltage-gated current sources. In the original model of Hodgkin and Huxley, the current sources are {\it memristive}, that is, they obey an Ohmic law but with conductances that have memory because they depend on the history of the voltage. In the present paper, such conductances with memory are called {\it memductances}, following the terminology of \citet{chuaMemristiveDevicesSystems1976}. The proposed standpoint is that this memristive modeling principle is valid both at the cellular and the population scale, and that only the model of the memductance differs. We then exploit the close analogy between ion channels (i.e.\ \emph{internal} memductances) and synapses (i.e.\ \emph{external} memductances) to model their temporal and spatial voltage dependence.

We stress the advantages of a simple circuit model that concentrates all the temporal and spatial modeling complexity in the spatiotemporal voltage dependence of the memductances. Simplified models preserve the same physical circuit structure but use simplified representations of the memductances. Temporal models spatially average the spatial dependence, while spatial models temporally average the voltage dependence. We argue that such models can be simulated, analyzed, and designed \emph{at scale} by modulating the model resolution of the memductances only. 

This paper is outlined as follows.
First, we introduce the memristive model of temporal and spatial memductances.
Then we exploit the simplicity and clean separability in space and time of the memductances to investigate the neural properties required for mean-field phenomena to remain robust to large parameter uncertainties.
This demonstrates a clear example of how the introduced model enables us to bridge properties from the neuronal microscale to the neural field macroscale,
and we discuss the implications for the tuning of behaviors of neural assemblies.
The final section concludes.

\section{The mixed-feedback motif of temporal excitability}
\label{sec:cellular-excitability}

The biophysical foundation of neuroscience,
and therefore of our understanding of excitable behaviors,
is the work of Hodgkin and Huxley,
culminating in \citet{hodgkinQuantitativeDescriptionMembrane1952}.

The key modeling principle is that a neuron can be represented by an RC circuit in parallel with a bank of voltage-gated current sources. The Hodgkin--Huxley model takes the form 
\begin{equation}
   \begin{aligned}
      C \dot{v}(t) &= - \left( i_l(t) + i_\text{Na}(t) + i_\text{K}(t) \right) + i_\text{app}(t) \\
      i_l(t) &= g_l \left( v(t) - E_l \right) \\ 
      i_\text{Na}(t) &= g_\text{Na}(t) \left( v(t) - E_\text{Na} \right) \\ 
      i_\text{K}(t) &= g_\text{K}(t) \left( v(t) - E_\text{K} \right), \\ 
   \end{aligned}
   \label{eq:HH-model}
\end{equation}
where $v$ is the voltage over the cell membrane and $C$ is the capacitance of the membrane. 
The circuit includes a sodium current $i_\text{Na}$, a potassium current $i_\text{K}$, and a leak current $i_l$ that lumps the contributions of all remaining ion channels together. The sodium current is excitatory (E) (or inward / negative) since the membrane potential is always {\it below} the Nernst potential $E_\text{Na}$. The potassium current is inhibitory (I) (or outward / positive) since the membrane potential is always {\it above} the Nernst potential $E_\text{K}$. 
The dynamics of the conductances are governed by activation variables, which increase conductances, and inactivation variables, which cause them to decrease.
Further details of the conductances $g_\text{Na}(t)$ and $g_\text{K}(t)$  are provided in Appendix \ref{methods:HH}. 
We call them {\it memductances} to emphasize the {\it memristive} nature of the Ohmic currents, namely, that the conductances have memory, that is, depend on the past history of the voltage. 
 
\begin{figure}
\centering
\resizebox{0.75\textwidth}{!}{
\begin{tikzpicture}[node distance=1.0cm]
	\node (v) at (-1,0) {$v(t)$};
	\node (start) [inner sep=0pt, outer sep=0pt] at (0,0) {};
    \node (minus_e) at ($(start)+(1.25cm:1cm)$) [round] {$+$};
    \node (times_e) at ($(minus_e)+(0:3cm)$) [node distance=0.5cm, minimum size=0.5cm, draw, inner sep=2pt] {$g_e(t)$};
    \node (minus_i) at ($(start)+(-1.25cm:1cm)$) [round] {$+$};
    \node (times_i) at ($(minus_i)+(0:3cm)$) [node distance=0.5cm, minimum size=0.5cm, draw, inner sep=2pt] {$g_i(t)$};
    \node (minus_currents) at ($(times_e)!0.5!(times_i)+(0:1cm)$) [round] {$+$};
    \node (negative_gain) at ($(minus_currents)+(0:1.5cm)$) [node distance=0.5cm, minimum size=0.5cm, draw, inner sep=2pt] {$-1$};
    \node (plus_currents) at ($(negative_gain)+(0:1cm)$) [round] {$+$};
    \node (RC) at ($(plus_currents)+(0:1.5cm)$) [block] {RC};
    \node (output) [right of=RC] {};

    \draw [-, thick] (v) -- (start);
    \draw [arrow] (start) |- (minus_e);
    \draw [arrow] (start) |- (minus_i);
    \draw[arrow] ([yshift=25]minus_e.north) -- node[near start, right] {$-E_e$} (minus_e.north);
    \draw[arrow] ([yshift=-25]minus_i.south) -- node[near start, right] {$-E_i$} (minus_i.south);
    \draw [arrow] (minus_e) -- node[above] {$v(t) - E_e$} (times_e);
    \draw [arrow] (minus_i) -- node[above] {$v(t) - E_i$} (times_i);
    \draw [arrow] (times_e) -| node[near end, right] {$i_e(t)$} (minus_currents);
    \draw [arrow] (times_i) -| node[near end, right] {$i_i(t)$} (minus_currents);
    \draw [arrow] (minus_currents) -- (negative_gain);
    \draw [arrow] (negative_gain) -- (plus_currents);
    \draw [arrow] ([yshift=30]plus_currents.north) -- node[near start, right] {$i_\text{app}(t)$} (plus_currents.north);
    \draw [arrow] (plus_currents) -- (RC);
    \draw [-, thick] (RC) -- (output);
    \draw [-, thick] (output.west)
    -- ++(0,-2cm)
    -| (v.east);
\end{tikzpicture}
}
\caption{Block diagram of the temporal mixed-feedback motif, as found, for example, in the Hodgkin--Huxley model, with $g_\text{Na} = g_e$ and $g_\text{K} = g_i$. 
Note that $i_e(t)$ is negative and $i_i(t)$ positive, so the currents have opposite signs.}
\label{fig:EI-block-diagram}
\end{figure}
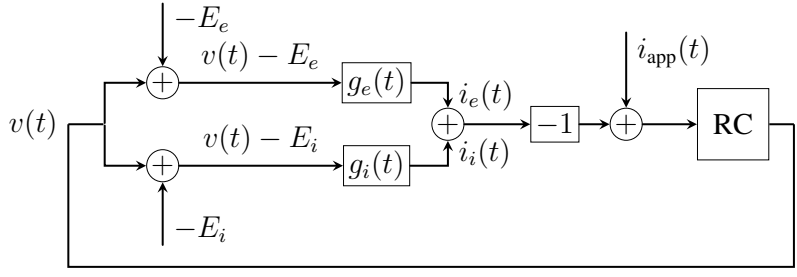

Figure \ref{fig:EI-block-diagram} highlights the mixed-feedback motif of the model, that is, the block diagram representation of an RC circuit modeling the passive membrane in feedback with two currents of opposite sign. Activation of the sodium current $i_\text{Na}(t)$ is a source of positive feedback, so it plays the role of the  excitatory current $i_e(t)$. It enables the destabilizing switch of the upward stroke, whereas activation of the potassium current $i_\text{K}$ (and inactivation of the sodium current) is a source of negative feedback, so it fulfills the role of $i_i(t)$, enabling the repolarization of the membrane after a refractory period. 

\begin{figure}
   \centering
   \includegraphics[width=0.7\textwidth]{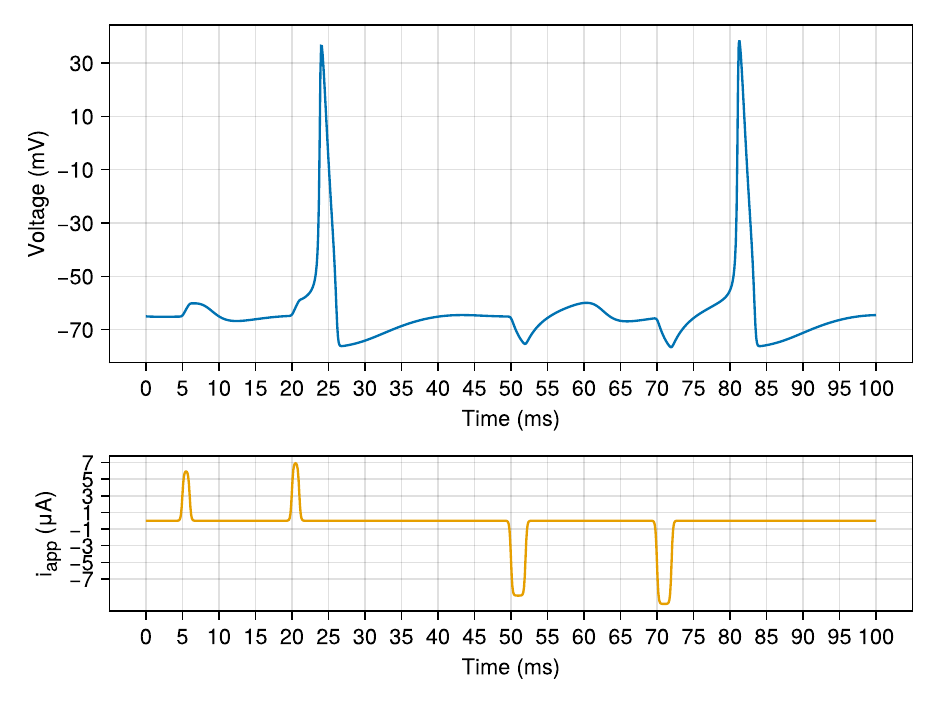}
   \caption{
      Temporal excitability in the Hodgkin--Huxley model.
      The threshold property is reflected in a small difference in $i_\text{app}$ (bottom) causing a large difference in the voltage response (top).
   }
   \label{fig:HH-excitability}
\end{figure}

We highlight that this mixed-feedback structure is reflected in two complementary properties of temporal excitability, illustrated in Figure~\ref{fig:HH-excitability}: a spike, or action potential, can be triggered either by a short-duration depolarizing current or upon release from a prolonged hyperpolarizing current. The latter phenomenon is known as \emph{rebound excitability}, or postinhibitory rebound. Rebound excitability plays an important role in the generation and modulation of neural rhythms \citep{dethierPositiveFeedbackCellular2015}, and has also been exploited as a computational mechanism for rhythmic generation in neuromorphic architectures \citep{huo2026}.

The coexistence of an excitatory and an  inhibitory threshold echoes the mixed-feedback structure of the model. Rebound excitability is lost in the absence of the potassium current. The temporal and amplitude properties of the two thresholds reflect the temporal and activation properties of the two currents.  

The complementarity of the two currents is also reflected in the current balance at equilibrium. The E and I currents are both active (non-zero) at equilibrium. Rebound excitability is a consequence of decreasing the (steady-state) activation of potassium and the (steady-state) inactivation of sodium when inhibiting the neuron,  leading to a rebound effect when released.

For design and analysis purposes, it is useful to simplify the details of the Hodgkin--Huxley model. Our proposal is to retain the mixed-feedback motif of Figure~\ref{fig:EI-block-diagram} and to retain the memristive nature of the feedback currents, but to simplify the modeling of the memductances. We choose to model a memductance $g_x$ as the series interconnection of a linear first-order lag with time constant $\tau_x$ and a sigmoidal nonlinearity: 
\begin{equation}
\begin{aligned}
    \tau_x \dot v_x(t) &= \left( v(t) - v_x(t) \right), \\
    x(t) &= \sigma_x(v_x(t)) := \sigma \left( \kappa_x \left( v_x(t) - v_x^{\text{th}} \right) \right),\\
    g_x(t) &= \bar{g}_x x(t)
\end{aligned}
\label{eq:canonical_temporal_memductance}
\end{equation}
where $v_x$ is the filtered voltage, $x \in [0, 1]$ is the activation variable which determines how open the channel is, and $\bar{g}_x$ is the maximal conductance.
The constant $\tau_x$ determines the timescale of activation, whereas the constants $\kappa_x$ and $v_x^{\text{th}}$ determine the voltage range of activation. To obtain a minimal model, we omit the inactivation of the sodium current, which is redundant with the activation of potassium. 

\begin{figure}
    \centering
    \includegraphics[width=0.7\linewidth]{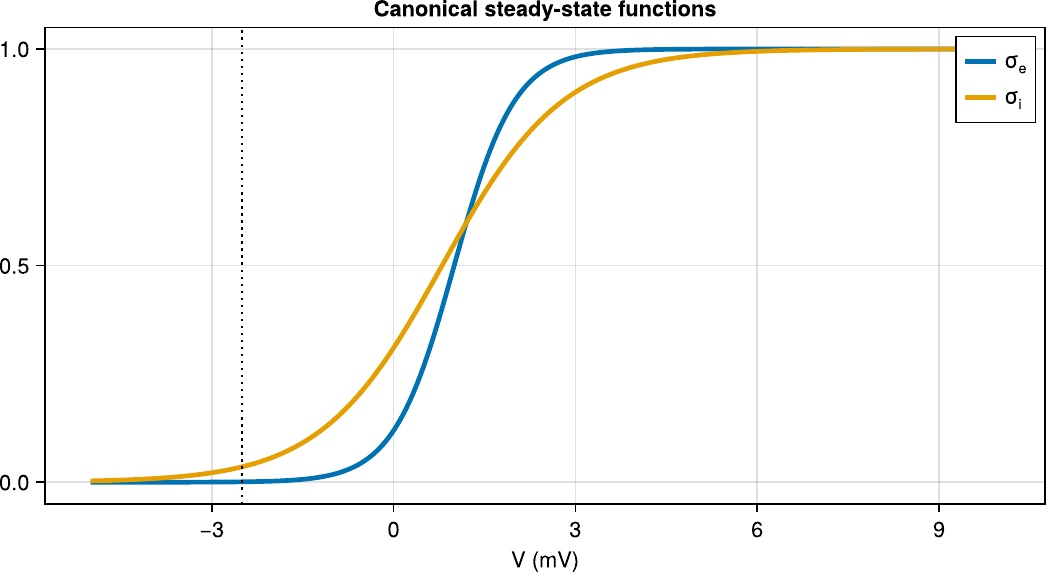}
    \caption{
        The nonlinear sigmoidal activation functions of the $e$ and $i$ memductances in the canonical model.
        The dotted line indicates the resting potential: note that the $i$ current in particular remains partially open at this $v_\text{rest}$.
    }
    \label{fig:canonical_activation_fns}
\end{figure}

We thus replace the sodium memductance $g_\text{Na}$ by a simplified excitatory (E) conductance $g_e$, and the potassium conductance $g_\text{K}$ by a simplified inhibitory (I) conductance $g_i$. The time constants are chosen so that the activation of the E current is {\it fast} relative to the {\it slow} activation of the $I$ current.  The steady-state activation functions of the EI model are shown in Figure~\ref{fig:canonical_activation_fns}.
The voltage dynamics become
\begin{equation}
    C \dot{v}(t) = - \left( i_e(t) + i_i(t) + i_l(t) \right) + i_\text{app}(t),
\end{equation}
which is the mixed-feedback motif from Figure~\ref{fig:EI-block-diagram},
with $i_{e / i}(t) = g_{e / i}(t) \left(v(t) - E_{e / i} \right)$, and $i_l(t) = \bar{g}_l \left( v(t) - E_l \right)$ being a purely resistive current.

\begin{figure}
   \centering
   \includegraphics[width=0.7\textwidth]{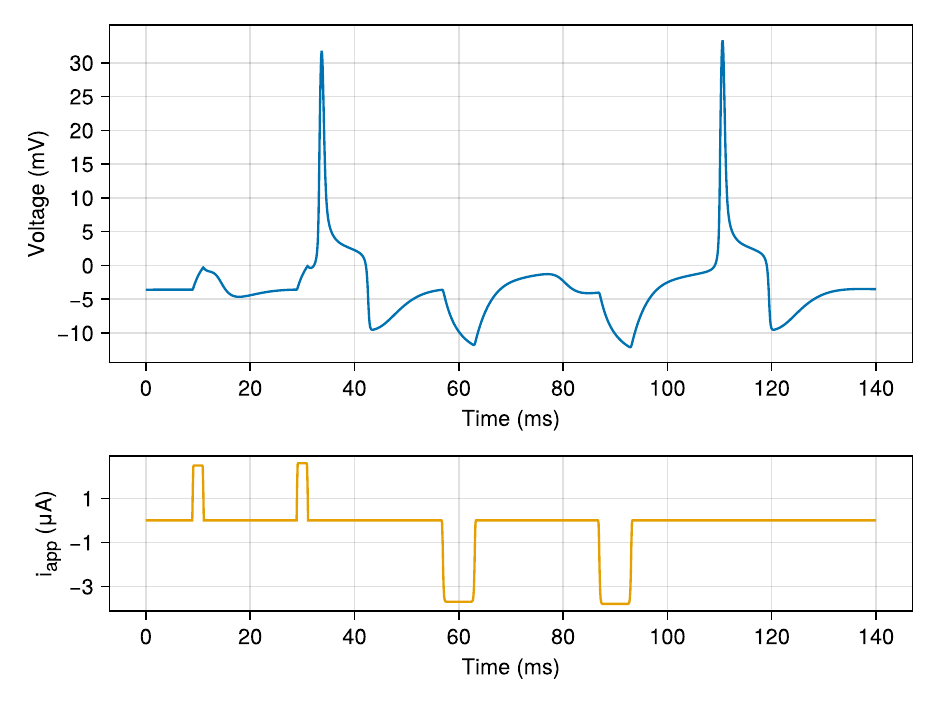}
   \caption{
      Temporal excitability of the memristive neuron model.
      A small increase in $i_\text{app}$ brings the voltage above threshold,
      resulting in a large pulse-like voltage trajectory (top).
      This threshold behavior is qualitatively equivalent to the excitability of the Hodgkin--Huxley model,
      cf.\ Figure \ref{fig:HH-excitability}.
   }
   \label{fig:canonical_excitability}
\end{figure}

Figure~\ref{fig:canonical_excitability} shows that the minimal EI model with the parameters shown in Table~\ref{tab:canonical-temporal-params} retains the two complementary threshold properties of the Hodgkin--Huxley model. Also in the EI model, rebound excitability is lost in the absence of the inhibitory current.

We emphasize that the excitable behavior of the EI model exists for large ranges of parameters. A simple analysis in Appendix~\ref{subsubsec:excitability_criterion} shows that the existence of both an excitatory and an inhibitory threshold requires only a sufficient timescale separation, that is, a sufficiently small ratio $\tau_e / \tau_i$, and activation functions that overlap as in Figure~\ref{fig:canonical_activation_fns}, that is, an inversion of the balance of the E and I activations near equilibrium.

The latter property encodes the criteria for having both an excitatory and an inhibitory threshold:
around and below the resting potential $\sigma_i(v) > \sigma_e(v)$, so inhibition can be removed by hyperpolarization,
but $\sigma_e'(v) > \sigma_i'(v)$ around the firing threshold,
so excitation can quickly dominate during rebound or depolarization.

The EI model is a  minimal model of excitability because it achieves both mechanisms without adding separate inactivation, deinactivation, or hyperpolarization-activated variables. 
It is robust because it does not rely on a finely tuned voltage point: 
hyperpolarization broadly reduces slow inhibition, while rebound or depolarization broadly enters the steep excitatory region. 
It also retains the interpretability of the Hodgkin--Huxley model because it mirrors the temporal logic of excitability: fast, sharp positive feedback from $e$, followed by slower, broader negative feedback from $i$.

\section{The mixed-feedback motif of lateral inhibition}
\label{sec:spatio-temporal-excitability}

Information in the brain is routed and shared through many synaptic stages and brain areas.
In the brain, the spatial extent of activity is mediated through a zoo of neurons,
which can be broadly characterized into two classes: excitatory (E) and inhibitory (I).
Excitatory neurons promote further neuronal activity,
while inhibitory neurons usually diminish it.
Excitation and inhibition are spatially structured,
with E and I operating in a balanced regime.
This can lead to interesting spatial patterns, such as oscillations, in the neuronal activity.
Two foundational models of such dynamics are those of \citet{wilsonExcitatoryInhibitoryInteractions1972} and \citet{amariDynamicsPatternFormation1977}.
This balance can manifest itself through lateral inhibition:
a circuit motif where active excitatory neurons suppress nearby neurons through inhibitory interneurons.

\begin{figure}
   \centering
   \includegraphics[width=0.7\textwidth]{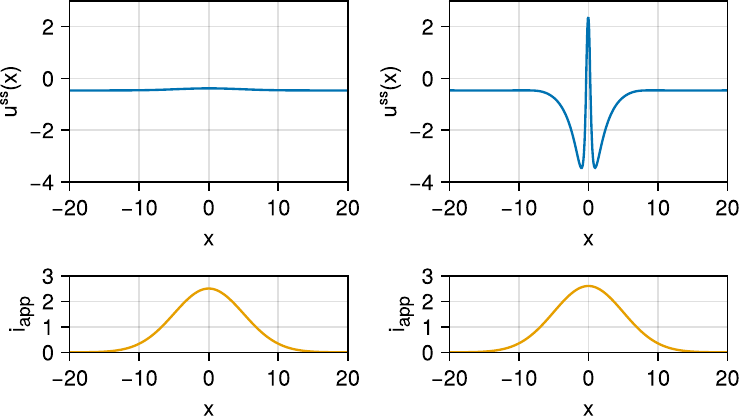}
   \caption{
      Spatial excitability from the mixed-feedback interaction kernel.
      Left: the subthreshold steady-state voltage response $u^{ss}(x)$ (top)
      to a Gaussian input current (bottom).
      Right: The superthreshold steady-state voltage response $u^{ss}(x)$ (top)
      to a Gaussian input current (bottom).
      The first applied pulse does not lead to spatial excitation,
      whereas the second applied pulse does.
   }
   \label{fig:spatial_excitability}
\end{figure}

Amari's model has the form
$$ C \frac{\partial v}{\partial t}(x,t) = - g_l \;  v(x,t) + (w * S(v))(x,t)$$
which we have chosen to write as a spatial analog of the Hodgkin--Huxley model (\ref{eq:HH-model}). 
Here, $S(v)$ is a firing-rate function and the notation $w*\cdot$ stands for spatial convolution with a kernel $w$ given by  
\begin{equation}
    w(x) = w_E e^{-|x| / \sigma_E} - w_I e^{-|x| / \sigma_I}, \quad \sigma_E < \sigma_I.
    \label{eq:amari-kernel}
\end{equation}
This model has the same mixed-feedback interpretation as in Figure~\ref{fig:EI-block-diagram}: a spatially extended RC circuit in feedback with two currents of opposite signs, having a short-range excitatory current, and a long-range inhibitory current. 
As in the temporal case, this spatial mixed-feedback motif gives rise to a threshold phenomenon:
when neural activity exceeds a threshold,
the feedback mechanism leads to a spatially localized `event' or `spike'.
The field organizes into active regions whose width is determined by the interaction kernel $w(x)$,
while activity elsewhere is suppressed.
This threshold behavior is illustrated in Figure~\ref{fig:spatial_excitability}.

\section{A mixed-feedback motif of\protect\\ spatiotemporal excitability}

Given the analogy between the motifs of temporal and spatial excitability, it seems only natural to mimic the same architecture in a spatiotemporal model of excitability. The memristive nature of currents is very useful to that end: the temporal or spatial kernel modulates the memductance of the Ohmic current, rather than the current itself.

To obtain a mixed-feedback motif, we separate the neural population into an excitatory and an inhibitory population,
and choose the spatial kernel $w(x) = e^{-|x| / \sigma}$.

 We regard the spatiotemporal memductances as modulating the {\it synaptic} currents, since they model interneuron coupling within the population:
\begin{equation}
\begin{aligned}
    \tau_\text{syn} \dot v_{\text{syn}}^{(T)}(x, t) &= v(x, t) - v_{\text{syn}}^{(T)}(x, t) \\
    v_{\text{syn}}^{(ST)}(x, t) &= \left( w * v_{\text{syn}}^{(T)} \right) (x,t) \\
    g_{\text{syn}}(x, t) &= \sigma_\text{syn} \left( v_\text{syn}^{(ST)}(x, t) \right) := \bar{g}_\text{syn} \sigma \left(\kappa_\text{syn} \left( v_{\text{syn}}^{(ST)}(x, t) - v_{\text{syn}}^{\text{th}} \right) \right),
\end{aligned}
\label{eq:canonical_spatial_memductance}
\end{equation}
where the superscript $(T)$ denotes a temporally filtered field,
the superscript $(ST)$ denotes the spatial filtering of the temporally filtered field,
$v^\text{th}_\text{syn}$ is the memductance's threshold,
$\kappa_\text{syn}$ determines its activation range,
and $*$ denotes spatial convolution.

The lateral inhibition profile of the spatial kernel requires $\frac{\sigma_E}{\sigma_I}$ to be small, in the same way as temporal excitability requires $\frac{\tau_{E}}{\tau_{I}}$ to be small. 

\begin{figure}
    \centering
    \resizebox{\textwidth}{!}{
\begin{tikzpicture}[
    block/.style={
        draw,
        minimum width=1.28cm,
        minimum height=0.62cm,
        align=center
    },
    sum/.style={
        draw,
        circle,
        minimum size=0.50cm,
        inner sep=0pt
    },
    line/.style={-Latex, thick},
    wire/.style={thick},
    labelnode/.style={font=\small, fill=white, inner sep=1.8pt},
    every node/.style={font=\small}
]

\def\xBus{0.00}
\def\xStub{0.0}

\def\xA{1.65-0.5} 
\def\xB{3.35}     
\def\xC{5.35}     
\def\xD{8.25}     

\def\xPair{10.45}
\def\xJoin{11.90}
\def\xNeg{12.70}  
\def\xFinal{13.80}
\def\xRC{15.20}
\def\xOut{16.25}

\def\yA{-0.90}
\def\yB{-2.55}
\def\yC{3.00}
\def\yD{1.35}

\coordinate (busMid) at (\xBus,0);
\draw[wire] (\xBus,3.00) -- (\xBus,-2.55);

\coordinate (inA) at (\xStub,\yA);
\coordinate (inB) at (\xStub,\yB);
\coordinate (inC) at (\xStub,\yC);
\coordinate (inD) at (\xStub,\yD);

\draw[wire] (\xBus,\yA) -- (inA);
\draw[wire] (\xBus,\yB) -- (inB);
\draw[wire] (\xBus,\yC) -- (inC);
\draw[wire] (\xBus,\yD) -- (inD);

\node[block] (b1a) at (\xA,\yA) {$*e^{-t / \tau_E}$};
\node[block] (b1b) at (\xB,\yA) {$*e^{- |x| / \sigma_E}$};
\node[block] (b1c) at (\xC,\yA) {$\sigma_E(\cdot)$};
\node[block] (b1d) at (\xD,\yA) {$v(t) - E_E$};

\draw[line] (inA) -- (b1a.west);
\draw[line] (b1a.east) -- (b1b.west);
\draw[line] (b1b.east) -- (b1c.west);
\draw[line] (b1c.east)
    -- node[labelnode, above=3pt] {$g_E(t)$}
    (b1d.west);

\node[block] (b2a) at (\xA,\yB) {$*e^{-t / \tau_I}$};
\node[block] (b2b) at (\xB,\yB) {$*e^{- |x| / \sigma_I}$};
\node[block] (b2c) at (\xC,\yB) {$\sigma_I(\cdot)$};
\node[block] (b2d) at (\xD,\yB) {$v(t) - E_I$};

\draw[line] (inB) -- (b2a.west);
\draw[line] (b2a.east) -- (b2b.west);
\draw[line] (b2b.east) -- (b2c.west);
\draw[line] (b2c.east)
    -- node[labelnode, above=3pt] {$g_I(t)$}
    (b2d.west);

\node[block] (b3a) at (\xA,\yC) {$*e^{-t / \tau_e}$};
\node[block] (b3c) at (\xC,\yC) {$\sigma_e(\cdot)$};
\node[block] (b3d) at (\xD,\yC) {$v(t) - E_e$};

\draw[line] (inC) -- (b3a.west);
\draw[line] (b3a.east) -- (b3c.west);
\draw[line] (b3c.east)
    -- node[labelnode, above=3pt] {$g_e(t)$}
    (b3d.west);

\node[block] (b4a) at (\xA,\yD) {$*e^{-t / \tau_i}$};
\node[block] (b4c) at (\xC,\yD) {$\sigma_i(\cdot)$};
\node[block] (b4d) at (\xD,\yD) {$v(t) - E_i$};

\draw[line] (inD) -- (b4a.west);
\draw[line] (b4a.east) -- (b4c.west);
\draw[line] (b4c.east)
    -- node[labelnode, above=3pt] {$g_i(t)$}
    (b4d.west);

\node[sum] (sumEI) at (\xPair,-1.72) {$+$};
\node[sum] (sumii) at (\xPair,2.18) {$+$};

\coordinate (eiTopIn) at (\xPair,\yA);
\coordinate (eiBotIn) at (\xPair,\yB);

\draw[line] (b1d.east)
    -- node[labelnode, above=3pt] {$i_E(t)$}
    (eiTopIn)
    -- (sumEI.north);

\draw[line] (b2d.east)
    -- node[labelnode, above=3pt] {$i_I(t)$}
    (eiBotIn)
    -- (sumEI.south);

\draw[line] (sumEI.east)
    -- node[labelnode, above=4pt] {$i_{\mathrm{syn}}(t)$}
    (\xJoin,-1.72);

\coordinate (iiTopIn) at (\xPair,\yC);
\coordinate (iiBotIn) at (\xPair,\yD);

\draw[line] (b3d.east)
    -- node[labelnode, above=3pt] {$i_e(t)$}
    (iiTopIn)
    -- (sumii.north);

\draw[line] (b4d.east)
    -- node[labelnode, above=3pt] {$i_i(t)$}
    (iiBotIn)
    -- (sumii.south);

\draw[line] (sumii.east)
    -- node[labelnode, below=4pt] {$i_{\mathrm{int}}(t)$}
    (\xJoin,2.18);


\node[
    block,
    minimum width=0.65cm,
    minimum height=0.52cm,
    inner sep=3pt
] (negGain) at (\xNeg,0.15) {$-1$};

\node[sum]   (sumFinal) at (\xFinal,0.15) {$+$};
\node[block] (RC)       at (\xRC,0.15) {RC};

\coordinate (Vout) at (\xOut,0.15);
\node[right=2pt] at ($(Vout)+(0,0.2)$) {$v(t)$};

\draw[wire] (\xJoin,2.18) -- (\xJoin,0.15);
\draw[wire] (\xJoin,-1.72) -- (\xJoin,0.15);

\draw[line] (\xJoin,0.15) -- (negGain.west);
\draw[line] (negGain.east) -- (sumFinal.west);

\draw[line] (\xFinal,2.50)
    -- node[labelnode, right=3pt] {$i_{\mathrm{app}}(t)$}
    (sumFinal.north);

\draw[line] (sumFinal.east) -- (RC.west);
\draw[line] (RC.east) -- (Vout);

\coordinate (Vtap) at ($(RC.east)!0.62!(Vout.west)$);

\draw[wire] (Vtap) -- ++(0.70,0)
    |- (0,-3.40)
    -| (busMid);

\end{tikzpicture}
}
\caption{
The spatiotemporal mixed-feedback structure of the canonical model.
The upper two branches represent the local temporal mixed-feedback structure, with a pointwise excitatory current $i_e$ and inhibitory current $i_i$, together making up the total internal current $i_\text{int}$.
The bottom two branches represent the interneuronal synaptic mixed-feedback structure, with an excitatory current $i_E$ and an inhibitory current $i_I$.
Together with an applied current $i_\text{app}$, these currents pass through an RC circuit to determine all the dynamics of the spatiotemporal network.
}
\label{fig:canonical_spatio_temporal_block_diagram}
\end{figure}
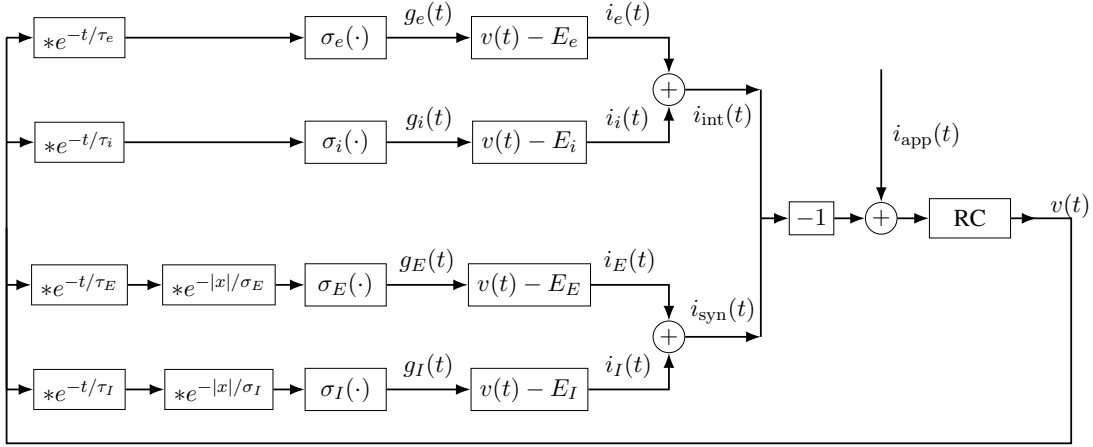

Combining this spatial mechanism with temporal excitability yields a model of spatiotemporal excitability,
the block diagram of which is drawn in Figure~\ref{fig:canonical_spatio_temporal_block_diagram}.
The model thus consists of the same  mixed-feedback motif as the temporal and spatial models of excitability. The only difference is that the currents  include both {\it internal} (that is, pointwise in space)  currents
and {\it synaptic} currents (that involve spatial coupling):
\begin{equation}
    C \dot v(x,t)
    =
    -\left(
        i_{\mathrm{int}}(x,t; v^x_t)
        +
        i_{\mathrm{syn}}(x,t; v^{\mathcal N(x)}_t)
    \right)
    + i_{\mathrm{app}}(x,t),
    \label{eq:CNN_v_dot}
\end{equation}
where the memductances of $i_\text{int}$ follow Equation~\ref{eq:canonical_temporal_memductance},
so they depend on the history of the voltage at $x$:
\begin{equation}
    v^x_t := \{v(x,s) : s \leq t\},
    \end{equation}
and those of $i_\text{syn}$ follow Equation~\ref{eq:canonical_spatial_memductance},
so they depend on the history of the voltage in a neighborhood around $x$, $\mathcal{N}(x)$:
\begin{equation}
    v^{\mathcal N(x)}_t := \{v(y,s) : y \in \mathcal N(x),\ s \leq t\}.
\end{equation}
From here on, we will denote spatial memductances with an uppercase $E / I$,
whereas internal, purely temporal memductances will be denoted with a lowercase $e / i$.

The architecture in Figure \ref{fig:canonical_spatio_temporal_block_diagram} therefore combines pointwise temporal mixed-feedback with spatially filtered excitatory–inhibitory coupling. 
The internal currents determine the local excitable dynamics at each spatial position, while the synaptic currents mediate recruitment and suppression across neighboring positions. 
The population implementation of this architecture and the collective event that it generates are considered in the following section.

It is worth emphasizing the simplicity of this model:
it consists of nothing more than a linear convolution in time followed by a nonlinearity for the `internal' currents,
and two linear convolutions (one in time and one in space) followed by a nonlinearity for the `synaptic' currents.
This simplicity makes its behavior straightforward to analyze and computationally inexpensive to simulate, allowing us to characterize the collective population response and subsequently investigate its robustness to parameter uncertainty.

\section{Population Excitability}

\subsection{Population spike}
\label{sec:population-spike}

The continuous spatiotemporal model introduced in the previous
section is simulated as two spatially aligned populations of
excitatory and inhibitory units. Each population contains \(N=901\)
units distributed uniformly over the spatial interval
\(x\in[-30,30]\). 
Each unit retains the local temporal mixed-feedback dynamics introduced in Section~\ref{sec:cellular-excitability}, 
while interactions between units are mediated by the spatially structured excitatory and inhibitory synaptic pathways introduced in Section~\ref{sec:spatio-temporal-excitability},
so that the dynamics of the E and I populations are given by Equation~\ref{eq:CNN_v_dot}.
Further details of the spatial discretization and numerical implementation are given in Appendix~\ref{methods:spatiotemporal_excitation}.

The units in the population are not assumed to be identical. To
represent cell-to-cell variability, the maximal memductances of the
internal currents are sampled according to
\begin{equation}
    \bar{g}
    =
    \bar{g}_0(1+\vartheta),
    \qquad
    \vartheta\sim\mathcal{U}(-0.4,0.4),
    \label{eq:population-heterogeneity}
\end{equation}
so that their values vary uniformly by \(\pm40\%\) around their
nominal values \(\bar{g}_0\). Consequently, any coherent population
response must emerge despite substantial variability in the local
excitable dynamics, rather than from the exact synchronization of
identical units.

We refer to the resulting collective event as a
\emph{population spike}. A population spike is a suprathreshold
response that is localized both spatially and temporally: a
contiguous subpopulation is recruited over a bounded region of
space, remains active during a finite time window, and subsequently
returns toward rest. The term therefore does not merely describe
several individual units crossing threshold at approximately the
same time. Rather, it denotes a coherent spatiotemporal event whose
extent and evolution are determined by interactions within the
population.

\begin{figure}
   \centering
   \includegraphics[width=0.8\textwidth]{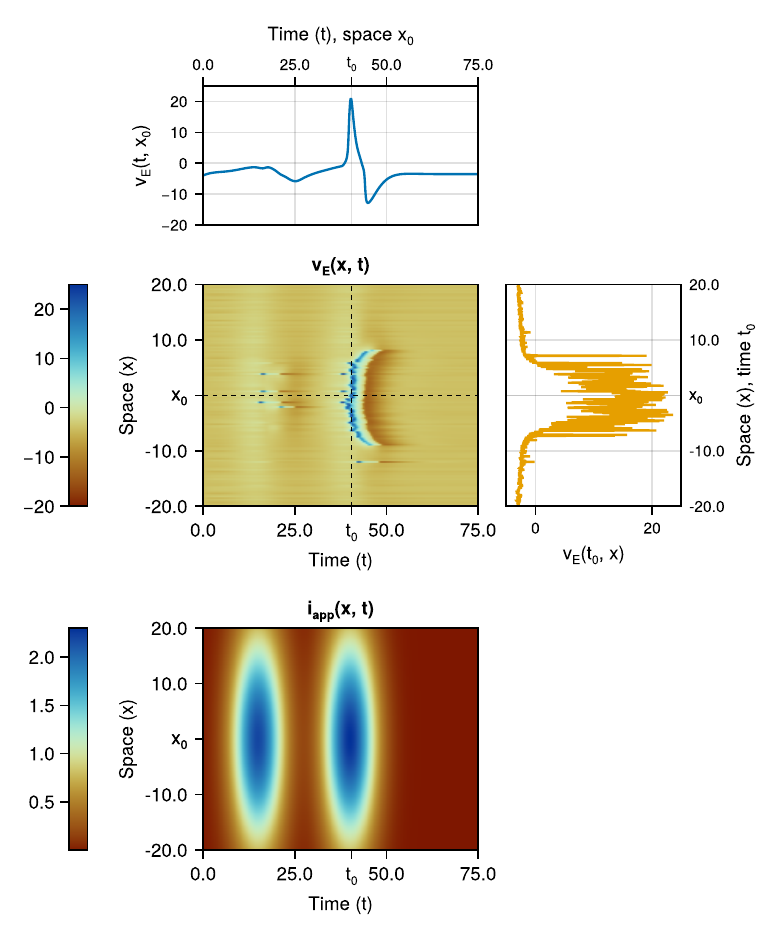}
\caption{
Population spike in the heterogeneous canonical model.
The central panel shows the excitatory population voltage
\(v_E(x,t)\) in response to the applied current
\(i_{\mathrm{app}}(x,t)\), shown in the lower panel. Despite
\(\pm40\%\) variability in the maximal internal memductances, a
coherent response emerges around \(x=x_0\) and \(t=t_0\). The upper
panel shows the temporal profile \(v_E(x_0,t)\), while the right-hand
panel shows the spatial profile \(v_E(x,t_0)\). Together, these
profiles demonstrate that the event is bounded in both time and
space: a localized subpopulation is recruited and subsequently
returns toward rest. We refer to this collective event as a
population spike.
}
   \label{fig:canonical_spatio-temporal_excitation}
\end{figure}

Figure~\ref{fig:canonical_spatio-temporal_excitation} demonstrates the formation of such a population spike in
the complete heterogeneous model. In response to the applied current,
the excitatory population voltage \(v_E(x,t)\) develops a pronounced
event centered at \(x=x_0\) and \(t=t_0\). The response recruits a
spatially bounded group of neighboring units and disappears after a
finite interval, demonstrating the simultaneous expression of
temporal and spatial excitability at the population level.

The temporal section through \(x=x_0\), shown in the upper panel of
Figure~\ref{fig:canonical_spatio-temporal_excitation}, displays the rapid voltage excursion and subsequent
recovery characteristic of temporal excitability. The spatial
section through \(t=t_0\), shown in the right-hand panel, shows that
this excursion is restricted to a finite region of the population.
Importantly, these profiles are not independent examples of temporal
and spatial excitability. They are two sections through the same
coherent event in the central space--time plot. Figure~\ref{fig:canonical_spatio-temporal_excitation} therefore
demonstrates a response that is localized simultaneously in space
and time.

The temporal and spatial scales of the population spike should be
distinguished from the corresponding scales of an individual
voltage spike. 
The spike of an individual neuron is determined
by its local temporal mixed feedback. 
By contrast, the duration and spatial extent of the population event additionally
depend on recruitment through excitation, suppression through
inhibition, the spatial ranges and timescales of the synaptic
kernels, and heterogeneity across the population. These are therefore
distinct model quantities, even when their numerical values happen
to be comparable. The population spike is not simply a larger copy
of an individual spike, nor merely the sum of many units spiking
simultaneously. It is a collective event with its own population-level
extent in space and time.

In Figure~\ref{fig:canonical_spatio-temporal_excitation}, the population event persists for approximately
\(T_{\mathrm{pop}}=\mathrm{3.3\,ms}\) and spans a spatial region of width
\(\ell_{\mathrm{pop}}=\mathrm{0.46}\), whereas the characteristic
duration of an individual voltage excursion is
\(T_{\mathrm{ind}}=\mathrm{1.3\,ms}\).
Thus, the population event persists for approximately 2.6 times as long as an individual excursion.
The temporal and spatial population widths are measured as the full widths at half maximum of the corresponding temporal and spatial population profiles, while \(T_{\mathrm{ind}}\) is the median full width at half maximum of the suprathreshold voltage excursions of individual (uncoupled) neurons.


Figure~\ref{fig:canonical_spatio-temporal_excitation} constitutes the central population-level result of the
present work. 
To the best of our knowledge, it provides the first
demonstration within a mixed-feedback framework that a
spatially coupled and strongly heterogeneous population of excitable
units can organize into a localized mean-field, or population-level,
spike. 
The essential result is that the individual units' activity is organized into a single coherent event with its own spatial and temporal extent,
rather than merely constituting a collection of individual voltage trajectories.

\subsection{Complementarity of temporal and spatial mixed-feedback}
\label{sec:population-ablation}

Having established that the complete heterogeneous model supports a
coherent population spike, we next ask which components of the
spatiotemporal architecture are required for its formation. In
particular, we investigate how the population response changes when
either spatial or temporal excitability is removed. These ablations
allow us to distinguish the collective event in Figure~\ref{fig:canonical_spatio-temporal_excitation} from
coincident responses caused only by a shared applied input.

\subsubsection{Complementarity of internal and synaptic currents}
\begin{figure}
    \centering
    \includegraphics[width=\columnwidth]{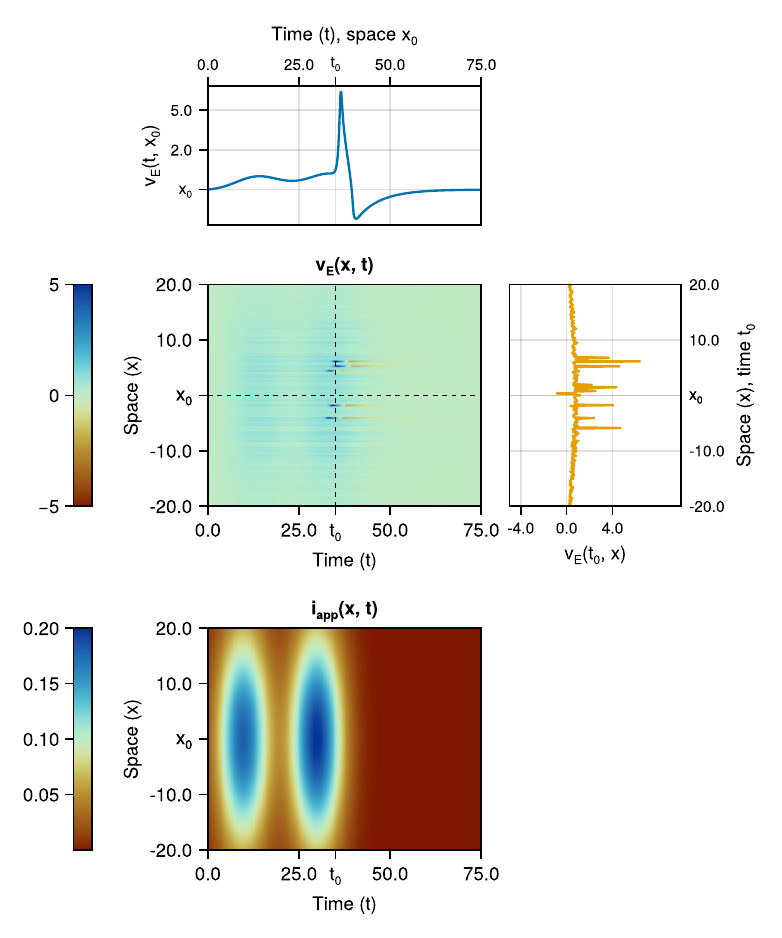}
    \caption{
    Spatial cohesion disappears from the neural population response without spatial memductances.
    The central top panel shows $v_E(x, t)$ in response to the applied current $i_\text{app}(x, t)$ shown below.
      The temporal profile of $v_E$ at $x_0 = 0$ is plotted in the upper inset,
      while the spatial profile of $v_E$ at $t_0 = 35$ is plotted in the upper-right inset.
      The spatial profile shows the absence of a spatially coherent response to the applied current.
    }
    \label{fig:no-spatial-temporal-excitability}
\end{figure}

We first remove the spatial memductances
\(g_{E}\) and \(g_{I}\) by setting
\(\bar{g}_{E/I}=0\), so that only the local internal
memductances \(g_{e/i}\) remain. In this regime, the units are
decoupled from the spatial mixed-feedback mechanism and the
population no longer displays coherent spatial activity
(Figure~\ref{fig:no-spatial-temporal-excitability}). 
Any apparent spatial agreement is accidental: 
it results from similar, but heterogeneous, excitable units responding independently to the same applied input.

\begin{figure}
    \centering
    \includegraphics[width=\columnwidth]{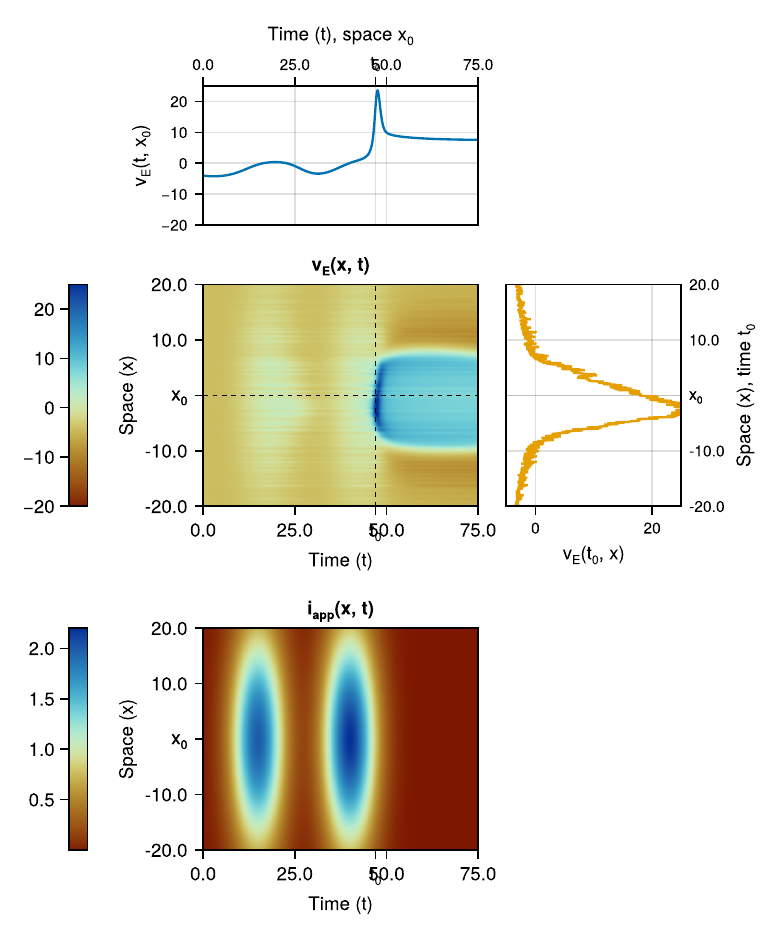}
    \caption{
    Cohesive temporal activity disappears in neural population response without temporal memductances.
      The response of the excitatory voltage $v_E(x, t)$ (central top figure)
      in response to an applied current $i_\text{app}(x, t)$ (bottom figure).
      The temporal profile of $v_E$ at $x_0 = 0$ is plotted in the upper inset,
      while the spatial profile of $v_E$ at $t_0 = 47$ is plotted in the upper-right inset.
      The temporal profile shows the absence of a temporally coherent response to the applied current.
    }
    \label{fig:no-temporal-spatial-excitability}
\end{figure}

We next examine the complementary case in which temporal
excitability is removed while spatial coupling is retained. 
To do so, we replace the dynamic internal memductances with an effective leak conductance and reversal potential chosen to preserve the resting current, 
and leave the spatial memductances in place. For
a fair comparison, the thresholds of the spatial memductances
\(v_{\mathrm{syn}}^{\mathrm{th}}\) are lowered to match those of the
internal memductances. As shown in Figure \ref{fig:no-temporal-spatial-excitability}, the resulting network
fails to display temporally coherent behaviors.
Even though the spatial memductances still provide positive feedback,
this feedback is very spatially localized due to the short-range excitatory kernel.
Due to parameter variability, the neighbors of a patch of active neurons need not cross their threshold,
and the active region is uneven.
Furthermore, an active patch need not necessarily return to rest,
as can be seen in Figure \ref{fig:no-temporal-spatial-excitability}.

\subsubsection{Complementarity of internal and interneuronal feedback}

\begin{figure}
    \centering
    \includegraphics[width=0.9\columnwidth]{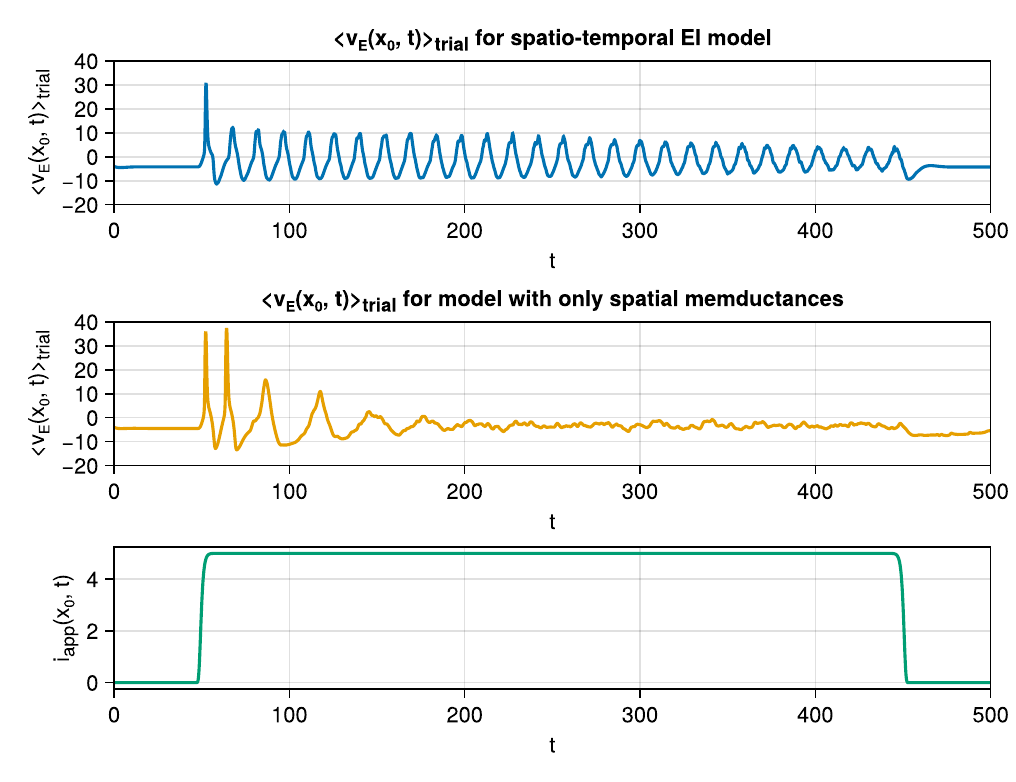}
    \caption{
    Temporal conductances help to maintain temporally cohesive activity.
    Average activity in a neural field at $x_0 = 0$ over $100$ trials in response to a long-lasting applied current (bottom)
    for a neural field with temporal memductances (top)
    and one without temporal memductances (middle).
    Note that the spatial memductances can facilitate an initial spiky response,
    but fail to sustain coherent temporal behaviors.
    }
    \label{fig:avg_activity_with_nontemporal_x0}
\end{figure}

To better illustrate the effect that the loss of the temporal memductances has on the temporal cohesion,
we applied a long-lasting input to the neural field with and without the temporal conductances for 100 trials, 
after which we computed the average voltage trajectory  at $(x_0, t)$ where $x_0 = 0$:
\begin{equation}
\langle v(x_0, t) \rangle_{\text{trial}} = \frac{1}{N_\text{trial}} \sum_{k=1}^{N_\text{trial}} v_k(x_0, t)
\end{equation}
The results are shown in Figure \ref{fig:avg_activity_with_nontemporal_x0}.
In the EI model, with both temporal and spatial memductances present, a neural rhythm is supported (Figure \ref{fig:avg_activity_with_nontemporal_x0}, top),
as would be expected for spatial $E / I$ coupling \citep{wilsonExcitatoryInhibitoryInteractions1972, borgersSynchronizationNetworksExcitatory2003}.
But in the model without the temporal memductances,
there is only a brief surge in activity due to the short-range excitatory connections, 
after which the spiking times become sufficiently desynchronized so that the trial-averaged voltage remains near the resting potential (Figure \ref{fig:avg_activity_with_nontemporal_x0}, middle).
This shows that, although initially the spatial memductances still allow a spike-like response to an input,
the behavior that immediately follows in time is not robust to parameter uncertainty.

This may seem surprising,
as the purely temporal and spatial memductances are equipped with the  same temporal convolution (cf.\ Equations \ref{eq:canonical_temporal_memductance} and \ref{eq:canonical_spatial_memductance}).
The critical difference is that for the temporal memductances the feedback loop is {\it internal}, making each individual neuron
endogenously excitable.
The  synaptic mixed feedback, in contrast, is external, that is, interneuronal.

\subsubsection{No mean-field excitability without both internal and synaptic currents}

\begin{figure}
   \centering
   \includegraphics[width=\columnwidth]{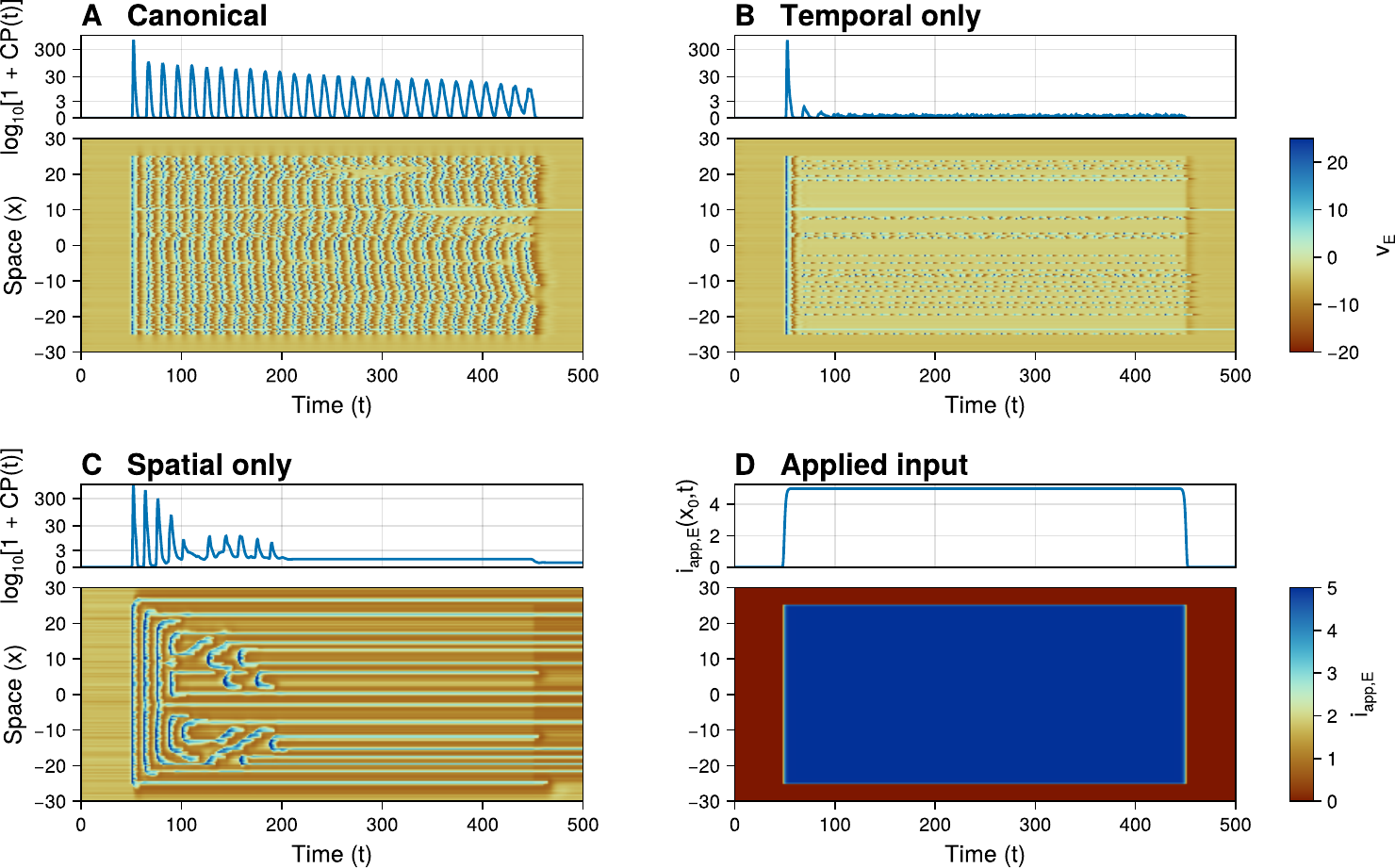}
\caption{
Robust spatiotemporal population rhythms in the EI model require both temporal and spatial memductances.
For each model condition in panels \textbf{A--C}, the upper trace shows the transformed coherent population activity
$\log_{10}\!\left[1+\mathrm{CP}(t)\right]$, where
$\mathrm{CP}(t)=\langle v_{\mathrm{E}}(x,t)\rangle_x^2$,
and the lower heatmap shows the excitatory voltage field
$v_{\mathrm{E}}(x,t)$.
The same realization of the heterogeneous parameters and initial conditions is
used in all three conditions, with independent pointwise variability of up to
$\pm40\%$ in the intrinsic conductance parameters
$g_l$, $\bar g_e$, and $\bar g_i$, and in the four synaptic conductance gains.
The voltage color scale is shared across panels \textbf{A--C}.
\textbf{(A)} Spatiotemporal EI model:  a robust sequence of population spikes whose amplitude gradually decays in time,
\textbf{(B)} EI model without spatial memductances,
\textbf{(C)} EI model without internal memductances,
\textbf{(D)} Applied current field $i_{\mathrm{app,E}}(x,t)$ (identical in all three experiments).
}
   \label{fig:PING_rhythm_comparison}
\end{figure}

The relevance of incorporating both temporal and spatial excitability can be seen more easily from a mean-field population event with scales beyond those of a single spatiotemporal spike.
Such events require many nearby neurons to be active simultaneously (so that the event happens in space),
and for them to be simultaneously active with high enough firing rates (so that the activity survives the temporal averaging).
This is illustrated in Figure \ref{fig:PING_rhythm_comparison}A:
it shows the activity in a coupled E--I network with both temporal and spatial memductances, as described above.

The activity heatmap shows the voltages of the neurons in the excitatory layer.
The network is heterogeneous, with the intrinsic conductance parameters \(g_l\), \(\bar g_e\), and \(\bar g_i\), together with the four synaptic conductance gains, varied independently by \(\pm40\%\) around their nominal values at each spatial position.
In Figure \ref{fig:PING_rhythm_comparison}A, top, 
we plot, as a measure of activity coherence, the square of the spatial average:
\begin{equation}
    \langle v \rangle_x^2(t) = \left( \frac{1}{x_\text{max} - x_\text{min}} \int_{x_\text{min}}^{x_\text{max}} v(x, t) \, dx \right)^2,
\end{equation}
where $x_\text{min}$ and $x_\text{max}$ are the spatial limits of the neural field.
This is a simple metric as it detects mean-field events in the population:
it can measure transient peaks in spatially coherent activity in the population,
which is precisely what a mean-field event is.

In the case where the model lacks either spatial memductances (Figure \ref{fig:PING_rhythm_comparison}B)
or temporal memductances (Figure \ref{fig:PING_rhythm_comparison}C),
no mean-field event can take place.
Without temporal memductances,
the spatial interactions within the neural field are no longer necessarily localized in time.
As a result, the neural field becomes sensitive to parameter heterogeneity,
and the activity is localized in space and fails to spread coherently, leading to spatial incoherence.
Without spatial memductances, the firing rates are high,
but the spike times of neighboring neurons are uncorrelated.
As a result, $\langle v \rangle_x^2 (t)$ settles on a baseline value and does not peak,
so no transient mean-field event takes place.

\subsection{Tuning through heterogeneity}
The robustness to heterogeneity described above opens up an avenue for tuning the population response of the neural field.
When changing parameters, such as the maximal memductances, of a subpopulation of neurons,
the qualitative population behavior is preserved: 
this is the robustness to heterogeneity illustrated above.
However, the neurons in this subset can then still be slower or faster than the average neuron in the population
and thereby act as an accelerator or brake on the whole, respectively.

\begin{figure}
   \centering
   \includegraphics[width=\columnwidth]{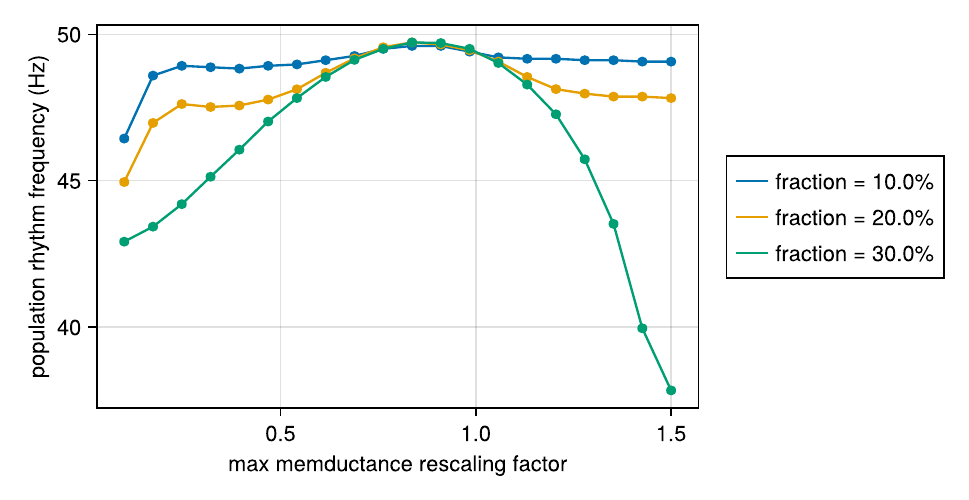}
   \caption{
      Tuning the population response by scaling the maximal memductances of a subset of the population.
      Here the population frequency is measured for a fixed input current $i_\text{app}$ 
      as the maximal memductances of internal currents of a central fraction of the population are rescaled.
   }
   \label{fig:heterogeneity_tuninig}
\end{figure}

To demonstrate this principle in the population rhythm from Figure \ref{fig:PING_rhythm_comparison}, 
we take a fraction of the population
and jointly rescale its intrinsic conductance parameters $g_l$, $\bar g_e$, and $\bar g_i$ by a factor $\mu > 0$: $\bar{g}_x^\text{new} = \mu \bar{g}_x$.
Heterogeneity in the parameters is preserved under this rescaling.
Then we measure how the frequency of the overall population oscillation is affected.
This is shown in Figure \ref{fig:heterogeneity_tuninig}.
Both the scaling factor and the fraction of the population that is affected by the scaling determine the magnitude of the frequency shift.
The upshot is that rescaling of the memductances in Figure \ref{fig:heterogeneity_tuninig} decreases the frequency of the population rhythm, by reducing the excitability of individual neurons. 

Tuning by heterogeneity is a simple mechanism. However, there is no such mechanism without robustness of the population to heterogeneity in the first place. It is because the mean-field population rhythm is robust to individual variability that the latter enables continuous modulation of parameters of the mean-field such as the mean-field frequency.

\section*{Conclusion}
The purpose of this paper was to introduce a model of excitability suitable for studying excitability across scales rather than at isolated temporal or spatial scales. 
Our starting point is the architecture of the Hodgkin--Huxley model: a capacitive membrane in feedback with two memristive currents of opposite signs. 
The proposed EI model captures excitability at any scale by incorporating memductances with dynamics at that scale. 
For this reason, it retains the biophysical interpretability of conductance-based modeling, whether at the cellular or population scale.

A central contribution of the EI model is that it separates the circuit architecture from the scale at which excitability is observed. The basic neuronal unit retains the familiar structure of cellular excitability, while spatiotemporal synaptic memductances retain the structure of lateral inhibition. The EI  model provides a unified methodological framework to analyze and design temporal excitability, spatial excitability, and their interaction in the generation of population excitability.

Because each conductance operates at a specific scale,
the neuronal behavior can be easily modified by changing the relative strength of the memductances at different scales.
Temporal excitability arises from fast positive feedback tempered by slower negative feedback on the membrane potential.
Spatial excitability is the result of short-range excitatory connections
coupled with longer-range inhibitory connections to limit the excitation and overall activity.
Together, these mechanisms give rise to population excitability: the tendency of spatially nearby neurons to produce temporally synchronized population events.

The key result of the paper is to show that both neuronal excitability and synaptic interconnections contribute to population excitability. The mean-field spike is not a mere copy of synchronized neuronal spikes. Instead, the population event has its own temporal and spatial scales, 
determined by the coherent activity of the individual neurons. 
Importantly, the mean-field spike is robust to a high level of heterogeneity in the population, modeled by a broad distribution of intrinsic and synaptic conductance parameters. 

Robustness to heterogeneity enables a key feature for design across scales: the heterogeneity of the population is the tunable knob of the mean-field. 
Such a mechanism could play a central role in the ability of neuromodulators to control excitability across a broad range of spatiotemporal scales. It is unclear how such mechanisms could unfold from 
classical neural-field models, \citep[e.g.,][]{amariDynamicsPatternFormation1977, wilsonExcitatoryInhibitoryInteractions1972}, or from the ubiquitous integrate-and-fire  models \citep[e.g.,][]{Lapicque1907}. 
From this perspective, we view the EI model as a candidate framework to bridge conductance-based models of cellular neurophysiology and mean-field models that can simulate entire populations. 

An important direction for future work is to connect the EI model more directly to electrophysiological and imaging data. 
Because the model expresses dynamical complexity through memductances, it may be possible to estimate these elements from measurements of cellular or population activity. 
Such estimates would make it possible to test whether the temporal and spatial feedback structures identified here are present in biological neural tissue, and whether their relative strengths predict the robustness or tunability of observed population events. 
This would also provide a route for comparing the EI model quantitatively with reduced spiking models, classical neural-field models, and more detailed conductance-based descriptions.

Taken together, these results suggest that neuronal activity across scales is shaped both by the modulation of excitability at the cellular level and by modulation of synaptic network topology.
By combining a physically interpretable conductance-based architecture with memductances that have explicit temporal and spatial dependence,
the EI model provides a single framework in which cellular mechanisms, parameter heterogeneity, and population dynamics can all be studied. 
Its main value is therefore not only that it reproduces particular neuronal behaviors, but that it clarifies which dynamical ingredients are needed for those behaviors to persist, vary, and organize across scales.

\subsection*{Acknowledgments}
This work was supported by funding from the European
Research Council under the European Union’s Horizon 2020
research and innovation program / ERC Advanced Grant:
SpikyControl (no. 7101054323).

\section*{Appendix A: Model and simulation details}
\label{sec:methods}

\setcounter{section}{1}
\setcounter{subsection}{0}
\setcounter{subsubsection}{0}

\renewcommand{\thesection}{A}
\renewcommand{\thesubsection}{A.\arabic{subsection}}
\renewcommand{\thesubsubsection}{\thesubsection.\arabic{subsubsection}}

\setcounter{equation}{0}
\renewcommand{\theequation}{A.\arabic{equation}}

\subsection{Hodgkin--Huxley model}
\label{methods:HH}

The sodium activation and inactivation variables, $m$ and $h$, and the potassium
activation variable, $n$, followed the standard Hodgkin--Huxley kinetics,
\begin{equation}
    \dot{x}
    =
    \alpha_x(v)(1-x)-\beta_x(v)x,
    \qquad
    x\in\{m,h,n\}.
\end{equation}
The voltage-dependent transition rates were
\begin{align}
    \alpha_m(v)
    &=
    \frac{0.1(v+40)}
    {1-\exp[-(v+40)/10]},
    &
    \beta_m(v)
    &=
    4\exp[-(v+65)/18],
    \\
    \alpha_h(v)
    &=
    0.07\exp[-(v+65)/20],
    &
    \beta_h(v)
    &=
    \frac{1}{1+\exp[-(v+35)/10]},
    \\
    \alpha_n(v)
    &=
    \frac{0.01(v+55)}
    {1-\exp[-(v+55)/10]},
    &
    \beta_n(v)
    &=
    0.125\exp[-(v+65)/80].
\end{align}

The Hodgkin--Huxley parameters used in the simulations are given in
Table~\ref{tab:hh-params}.

\begin{table}[h]
\renewcommand{\arraystretch}{1.3}
\caption{Parameters used for the Hodgkin--Huxley neuron model.}
\label{tab:hh-params}
\begin{center}
\begin{tabular}{|c|c|c|c|c|c|c|}
\hline
    $C$ &
    $\bar{g}_{\mathrm{Na}}$ &
    $E_{\mathrm{Na}}$ &
    $\bar{g}_{\mathrm{K}}$ &
    $E_{\mathrm{K}}$ &
    $\bar{g}_{\mathrm{L}}$ &
    $E_{\mathrm{L}}$
    \\
\hline
    $1$ &
    $120$ &
    $50$ &
    $36$ &
    $-77$ &
    $0.3$ &
    $-54.4$
    \\
\hline
\end{tabular}
\end{center}
\end{table}

Here, $C$ is measured in $\mu\mathrm{F}\,\mathrm{cm}^{-2}$,
the maximal conductances in $\mathrm{mS}\,\mathrm{cm}^{-2}$,
and the reversal potentials in $\mathrm{mV}$.

For Figure~\ref{fig:HH-excitability}, the applied current was constructed
from smooth rectangular pulses. Defining
$S(t;t_0,k)=[1+\exp(-k(t-t_0))]^{-1}$ and
$P(t;A,t_a,t_b)=A[S(t;t_a,k)-S(t;t_b,k)]$, with
$k=10~\mathrm{ms}^{-1}$, the applied current was
\begin{equation}
    I_{\mathrm{app}}(t)
    =
    P(t;6,5,6)
    +P(t;7,20,21)
    +P(t;-9,50,52)
    +P(t;-10,70,72).
\end{equation}
Pulse amplitudes are measured in $\mu\mathrm{A}\,\mathrm{cm}^{-2}$ and times
in $\mathrm{ms}$.

The system was initialized at
$v(0)=-65~\mathrm{mV}$, $m(0)=0.05$, $h(0)=0.6$, and $n(0)=0.32$,
and integrated over $0\leq t\leq100~\mathrm{ms}$ using the adaptive
fifth-order Tsitouras Runge--Kutta method (\texttt{Tsit5}), with relative
and absolute tolerances of $10^{-6}$. The numerical solution was sampled
at intervals of $\Delta t=0.1~\mathrm{ms}$.

\subsection{Amari's model}

For the Amari neural field simulations of
Figure~\ref{fig:spatial_excitability}, the firing-rate function was taken
to be a smooth approximation of a Heaviside threshold,
\begin{equation}
    S(v)
    =
    \frac{1}{1+\exp[-5(v-\theta)]},
    \qquad
    \theta=0.4.
\end{equation}
Spatial coupling was described by the Mexican-hat kernel
\begin{equation}
    w(x)
    =
    A\exp\left(-\frac{|x|}{a}\right)
    -
    B\exp\left(-\frac{|x|}{b}\right),
\end{equation}
with $A=2$, $B=1$, $a=0.3$, and $b=2$. The field time constant was
$\tau=3$.

The external input consisted of two spatially localized Gaussian
perturbations,
\begin{equation}
    I_{\mathrm{app}}(x,t)
    =
    \sum_{j=1}^{2}
    A_j
    \exp\left[
        -\frac{x^2}{2\sigma_x^2}
        -\frac{(t-t_j)^2}{2\sigma_t^2}
    \right],
\end{equation}
with $(A_1,A_2)=(2.5,2.6)$, $(t_1,t_2)=(10,210)$, and
$\sigma_x=\sigma_t=5$.

The spatial domain $x\in[-20,20]$ was discretized using $501$ uniformly
spaced points. Periodic boundary conditions were imposed, and the spatial
interaction term was evaluated as a discrete circular convolution using
the fast Fourier transform. The field was initialized at $v(x,0)=0$ and
integrated over $0\leq t\leq400$ using \texttt{Tsit5}, with relative and
absolute tolerances of $10^{-6}$ and a sampling interval of $\Delta t=0.5$.

\subsection{EI model}

The intrinsic conductance parameters of the EI model are given in
Table~\ref{tab:canonical-temporal-params}.

\begin{table}[h]
\renewcommand{\arraystretch}{1.3}
\caption{Temporal parameters used in simulations of the EI model in this paper.}
\label{tab:canonical-temporal-params}
\begin{center}
\begin{tabular}{|c|c|c|c|c|c|c|c|c|c|c|c|}
\hline
$g_l$ &
$E_l$ &
$\tau_e$ &
$\bar{g}_e$ &
$v_e^{\mathrm{th}}$ &
$\kappa_e$ &
$E_e$ &
$\tau_i$ &
$\bar{g}_i$ &
$v_i^{\mathrm{th}}$ &
$\kappa_i$ &
$E_i$
\\
\hline
$0.3$ &
$-1$ &
$1$ &
$3$ &
$1$ &
$2$ &
$50$ &
$10$ &
$20$ &
$1.2$ &
$1$ &
$-15$
\\
\hline
\end{tabular}
\end{center}
\end{table}

Unless otherwise stated, simulations of the EI model were integrated using
\texttt{Tsit5} with relative and absolute tolerances of $10^{-5}$.

\subsubsection{EI model excitability criterion}
\label{subsubsec:excitability_criterion}

We can approximately compute when the EI model will exhibit excitable behaviors.
Consider the model starting from rest with
$v = v_{e} = v_{i} = 0$ and $i_{\text{app}}$ being applied at $t=0$
with sufficient energy to push $v$ into the activation range of the
excitatory memductance. To obtain fast positive feedback, we require
$d i_{\text{tot}}/d v < 0$. Since $\tau_{e} < \tau_{i}$, we have
$v_{e} \approx v$, while $v_{i}$ remains close to rest. If
$\sigma_i(0)$ is small, then
\[
   i_{\text{tot}}
   \approx
   g_l v
   +
   \bar{g}_e \sigma_e(v)(v - E_e),
\]
so
\[
   \frac{d i_{\text{tot}}}{d v}
   \approx
   g_l
   +
   \bar{g}_e
   \left[
      \sigma_e(v)
      +
      (v - E_e)\sigma_e'(v)
   \right].
\]
The criterion for positive feedback is therefore
\[
   g_l
   +
   \bar{g}_e
   \left[
      \sigma_e(v)
      +
      (v - E_e)\sigma_e'(v)
   \right]
   < 0,
\]
or equivalently
\begin{equation}
   \sigma_e'(v)(E_e - v)
   >
   \sigma_e(v)
   +
   \frac{g_l}{\bar{g}_e}.
   \label{eq:pos_feedback_criterion}
\end{equation}
Thus, the excitatory reversal potential must lie above the voltage during
the rising phase, and the excitatory sigmoid must be sufficiently steep
and sufficiently strongly coupled to overcome the leak.

When both the excitatory and inhibitory conductances have activated, the
total current is approximately
\[
   i_{\text{tot}}
   \approx
   g_l v
   +
   \bar{g}_e \sigma_e(v)(v - E_e)
   +
   \bar{g}_i \sigma_i(v)(v - E_i).
\]
The differential conductance is then
\[
   \frac{d i_{\text{tot}}}{d v}
   =
   g_l
   +
   \bar{g}_e
   \left[
      \sigma_e(v)
      +
      (v - E_e)\sigma_e'(v)
   \right]
   +
   \bar{g}_i
   \left[
      \sigma_i(v)
      +
      (v - E_i)\sigma_i'(v)
   \right].
\]
The criterion for recovery, or slow negative feedback, is
\[
   g_l
   +
   \bar{g}_e
   \left[
      \sigma_e(v)
      +
      (v - E_e)\sigma_e'(v)
   \right]
   +
   \bar{g}_i
   \left[
      \sigma_i(v)
      +
      (v - E_i)\sigma_i'(v)
   \right]
   > 0.
\]
Since $E_i < 0$, the inhibitory contribution is positive for depolarized
voltages and can restore positive differential conductance once the slow
inhibitory activation catches up.

We can also determine the conditions for inhibitory-threshold excitation,
or rebound excitation after release from hyperpolarization. Consider the
model initially at rest and apply a hyperpolarizing current
$i_{\mathrm{app}}<0$, which drives the voltage to a negative value
$v_b$. At the time of release, $i_{\mathrm{app}}$ is set back to zero.
Let the state at release be
\[
   v(0)=v_b,
   \qquad
   v_e(0)=v_{e,b},
   \qquad
   v_i(0)=v_{i,b}.
\]
A necessary condition for rebound is that the voltage initially moves
upward after release. With
\[
   C\dot v = -i_{\mathrm{tot}} + i_{\mathrm{app}},
\]
this requires
\[
   i_{\mathrm{tot}}(v_b,v_{e,b},v_{i,b}) < 0,
\]
or
\[
   g_l v_b
   +
   \bar g_e \sigma_e(v_{e,b})(v_b-E_e)
   +
   \bar g_i \sigma_i(v_{i,b})(v_b-E_i)
   <0.
\]
This condition gives passive relaxation toward rest. For excitation,
however, the rebound trajectory must also enter a region of negative
differential conductance.

Immediately after release, $\tau_e<\tau_i$ implies that $v_e$ tracks the
recovering voltage, whereas $v_i$ remains close to its hyperpolarized
value. Thus
\[
   v_e \approx v,
   \qquad
   v_i \approx v_{i,b}.
\]
During this early rebound phase,
\[
   i_{\mathrm{tot}}
   \approx
   g_l v
   +
   \bar g_e \sigma_e(v)(v-E_e)
   +
   \bar g_i \sigma_i(v_{i,b})(v-E_i).
\]
Define the early-rebound differential conductance
\[
   D_{\mathrm{rise}}(v;v_{i,b})
   =
   g_l
   +
   \bar g_e
   \left[
      \sigma_e(v)
      +
      (v-E_e)\sigma_e'(v)
   \right]
   +
   \bar g_i \sigma_i(v_{i,b}).
\]
Then rebound positive feedback requires
\[
   D_{\mathrm{rise}}(v;v_{i,b}) < 0.
\]
Using
\[
   \sigma_e'(v)
   =
   \kappa_e
   \sigma_e(v)\left(1-\sigma_e(v)\right),
\]
this condition is
\[
   g_l
   +
   \bar g_e
   \left[
      \sigma_e(v)
      +
      \kappa_e (v-E_e)
      \sigma_e(v)\left(1-\sigma_e(v)\right)
   \right]
   +
   \bar g_i \sigma_i(v_{i,b})
   <0,
\]
or equivalently
\begin{equation}
   \kappa_e (E_e-v)
   \sigma_e(v)\left(1-\sigma_e(v)\right)
   >
   \sigma_e(v)
   +
   \frac{g_l+\bar g_i\sigma_i(v_{i,b})}{\bar g_e}.
   \label{eq:rebound_pos_feedback_criterion}
\end{equation}
Thus, rebound excitation requires the hyperpolarizing pulse to deactivate
the slow inhibitory conductance sufficiently, so that
$\sigma_i(v_{i,b})$ is small. Upon release, the voltage passively rebounds
toward rest; if it reaches a voltage range where the fast excitatory
sigmoid is sufficiently steep, the excitatory current creates negative
differential conductance and the rebound is amplified into a spike.

Recovery requires that the slow inhibitory conductance catches up during
the depolarized phase. Once both conductances track the voltage,
\[
   v_e\approx v,
   \qquad
   v_i\approx v.
\]
Define the recovery differential conductance
\[
   D_{\mathrm{rec}}(v)
   =
   g_l
   +
   \bar g_e
   \left[
      \sigma_e(v)
      +
      (v-E_e)\sigma_e'(v)
   \right]
   +
   \bar g_i
   \left[
      \sigma_i(v)
      +
      (v-E_i)\sigma_i'(v)
   \right].
\]
Then recovery requires
\[
   D_{\mathrm{rec}}(v)>0.
\]
Substituting the sigmoid derivatives gives
\[
   D_{\mathrm{rec}}(v)
   =
   g_l
   +
   \bar g_e
   \left[
      \sigma_e(v)
      +
      \kappa_e (v-E_e)
      \sigma_e(v)\left(1-\sigma_e(v)\right)
   \right]
   +
   \bar g_i
   \left[
      \sigma_i(v)
      +
      \kappa_i (v-E_i)
      \sigma_i(v)\left(1-\sigma_i(v)\right)
   \right].
\]
Equivalently, recovery requires
\begin{equation}
   \kappa_e (E_e-v)
   \sigma_e(v)\left(1-\sigma_e(v)\right)
   <
   \sigma_e(v)
   +
   \frac{
      g_l
      +
      \bar g_i
      \left[
         \sigma_i(v)
         +
         \kappa_i (v-E_i)
         \sigma_i(v)\left(1-\sigma_i(v)\right)
      \right]
   }{\bar g_e}.
   \label{eq:recovery_criterion}
\end{equation}

In summary, rebound excitation followed by recovery requires a sign change
in the effective differential conductance:
\begin{equation}
   D_{\mathrm{rise}}(v_{\mathrm{rise}};v_{i,b})
   <
   0
   <
   D_{\mathrm{rec}}(v_{\mathrm{rec}}).
   \label{eq:rebound_sign_change_criterion}
\end{equation}
Here $v_{\mathrm{rise}}$ denotes a voltage reached during early rebound,
when $v_e\approx v$ and $v_i\approx v_{i,b}$, whereas
$v_{\mathrm{rec}}$ denotes a later depolarized voltage, when
$v_e\approx v_i\approx v$.

In summary, rebound excitation is favored when $\sigma_i (v_{i,b}) \ll 1$.
In particular, $g_i(v_\mathrm{rest}) \not\approx 0$, that is, the inhibitory conductance must already be active at equilibrium.
Around and below the resting potential we require $\sigma_i(v) > \sigma_e(v)$, so the inhibition can be removed by hyperpolarization,
but around the firing threshold we require $\sigma'_e(v) > \sigma'_i(v)$, so the excitation can quickly come to dominate the dynamics when the firing threshold is crossed.

\subsubsection{Excitability simulations}
\label{methods:spatiotemporal_excitation}

For convenience, we define the super-Gaussian function
\begin{equation}
    G_p(z;z_0,\sigma)
    =
    \exp\left[
        -\frac{1}{2}
        \left(
            \frac{|z-z_0|}{\sigma}
        \right)^p
    \right].
\end{equation}

For Figure~\ref{fig:canonical_excitability}, synaptic interactions were
omitted and the neuron was driven by
\begin{equation}
    I_{\mathrm{app}}(t)
    =
    \sum_{j=1}^{4}
    A_j G_{30}(t;t_j,\sigma_j),
\end{equation}
where $(A_j)=(2.6,2.7,-3.6,-3.7)$,
$(t_j)=(10,30,60,90)$, and $(\sigma_j)=(1,1,3,3)$.
Thus, two brief depolarizing pulses were followed by two longer
hyperpolarizing pulses. The voltage and intrinsic filtered-voltage
variables were initialized at
$v(0)=v_e(0)=v_i(0)=-3.6$.
The system was integrated over $0\leq t\leq140$ and sampled at
intervals of $\Delta t=0.1$.

The spatial and synaptic parameters used for the spatiotemporal
excitability simulations are given in
Table~\ref{tab:spatial-synaptic-params}.

\begin{table}[h]
\renewcommand{\arraystretch}{1.3}
\caption{Spatial and synaptic parameters used in the spatiotemporal excitability simulations.}
\label{tab:spatial-synaptic-params}
\begin{center}
\begin{tabular}{|c|c|c|c|c|c|c|c|c|c|c|c|}
\hline
$\sigma_E$ &
$\sigma_I$ &
$\bar{g}^{\mathrm{syn}}_E$ &
$\bar{g}^{\mathrm{syn}}_I$ &
$E_E^{\mathrm{syn}}$ &
$E_I^{\mathrm{syn}}$ &
$v_E^{\mathrm{th}}$ &
$\kappa_E$ &
$v_I^{\mathrm{th}}$ &
$\kappa_I$ &
$\tau_E$ &
$\tau_I$
\\
\hline
$1$ &
$3$ &
$1.2$ &
$2.0$ &
$50$ &
$-15$ &
$5$ &
$1$ &
$0$ &
$2$ &
$2$ &
$10$
\\
\hline
\end{tabular}
\end{center}
\end{table}

The two fields were simulated on $x\in[-30,30]$ using 901 uniformly spaced
grid points. Self-coupling was removed from both spatial kernels, after which
each kernel was renormalized to unit spatial integral. Unless otherwise stated,
spatial heterogeneity was introduced independently at each grid point by
varying $g_l$, $\bar g_e$, and $\bar g_i$ uniformly within $\pm40\%$ of their
nominal values, while the synaptic maximal conductances were spatially
homogeneous.

For Figures~\ref{fig:canonical_spatio-temporal_excitation},
\ref{fig:no-spatial-temporal-excitability}, and
\ref{fig:no-temporal-spatial-excitability}, the excitatory field was driven by
\begin{equation}
    I_{\mathrm{app}}(x,t)
    =
    \sum_{j=1}^{2}
    A_j
    G_2(x;0,14)
    G_2(t;t_j,5),
\end{equation}
with $(A_1,A_2)=(2.0,2.2)$ and $(t_1,t_2)=(15,40)$.
The inhibitory field received no direct applied current.

The initial voltages were drawn independently according to
\begin{equation}
    v_\alpha(x,0)
    =
    -4+0.05\,\xi_\alpha(x),
    \qquad
    \alpha\in\{E,I\},
    \qquad
    \xi_\alpha(x)\sim\mathcal{N}(0,1),
\end{equation}
with the corresponding filtered variables initialized to the same values.
The system was integrated over $0\leq t\leq75$ and sampled at intervals
of $\Delta t=0.1$.

In Figure~\ref{fig:canonical_spatio-temporal_excitation}, the excitatory
voltage field $v_E(x,t)$ and applied current $I_{\mathrm{app}}(x,t)$ were
shown together with temporal and spatial projections at $x_0=0$ and
$t_0=40.5$.

For Figure~\ref{fig:no-spatial-temporal-excitability}, the spatial
interactions were removed by setting the excitatory and inhibitory
synaptic conductances to zero; the spatial projection was taken at
$t_0=35$. For Figure~\ref{fig:no-temporal-spatial-excitability}, the
dynamic intrinsic conductances were replaced by the effective leak
conductance
\begin{equation}
    g_l^{\mathrm{eff}}(x)
    =
    g_l(x)
    +
    g_e(x;v_{\mathrm{rest}})
    +
    g_i(x;v_{\mathrm{rest}}),
    \qquad
    v_{\mathrm{rest}}=-3.8,
\end{equation}
with an effective reversal potential chosen to preserve the total resting
intrinsic current. Synaptic coupling was retained, and the spatial
projection was taken at $t_0=47$.

For Figure~\ref{fig:avg_activity_with_nontemporal_x0}, $100$ paired trials
were simulated. In each trial, $g_l$, $\bar{g}_e$, and $\bar{g}_i$,
together with the baseline synaptic gains, were sampled independently at
each spatial grid point from a uniform $\pm40\%$ variation about their
nominal values. The same sampled parameter fields and initial conditions
were used for the two paired models. Initial voltages were drawn as
\begin{equation}
    v_\alpha(x,0)
    =
    -4+0.1\,\xi_\alpha(x),
    \qquad
    \alpha\in\{E,I\},
    \qquad
    \xi_\alpha(x)\sim\mathcal{N}(0,1),
\end{equation}
with independent standard normal variates $\xi_\alpha(x)$.

In the purely synaptic model, the dynamic intrinsic excitatory and
inhibitory conductances were removed and replaced by spatial synaptic
channels matched to the corresponding intrinsic channels, while the leak
and baseline synaptic network were retained. Both models were driven by
the same broad super-Gaussian input,
\begin{equation}
    I_{\mathrm{app}}(x,t)
    =
    5G_{200}(x;0,25)G_{200}(t;250,200),
    \label{eq:broad_supgauss}
\end{equation}
and simulated over $0\leq t\leq500$ with a sampling interval of
$\Delta t=0.1$. The quantity shown in
Figure~\ref{fig:avg_activity_with_nontemporal_x0} is the excitatory voltage
at $x_0=0$, averaged pointwise over trials,
\begin{equation}
    \left\langle
        v_E(x_0,t)
    \right\rangle_{\mathrm{trial}},
\end{equation}
shown separately for the canonical and purely synaptic models together
with $I_{\mathrm{app}}(x_0,t)$.

For Figure~\ref{fig:PING_rhythm_comparison}, a single realization of the
heterogeneous population was simulated. The intrinsic conductance parameters
$g_l$, $\bar g_e$, and $\bar g_i$, together with the four baseline synaptic
gains, were varied independently by $\pm40\%$ around their nominal values at
each spatial position. The same parameter realization and initial conditions
were used in all three conditions. Initial voltages were sampled independently
from $\mathcal{N}(-4,0.1^2)$, with all filtered-voltage variables initialized
to the corresponding membrane voltages. The excitatory population was driven
by the broad super-Gaussian input of Equation~\ref{eq:broad_supgauss}, while
the inhibitory population received no direct input. The system was simulated
over $0 \leq t \leq 500$ and sampled at intervals of $\Delta t=0.2$.

Three conditions were compared: the canonical model; a temporal-only model,
in which all spatial synaptic conductances were set to zero; and a spatial-only
model, in which the dynamic intrinsic excitatory and inhibitory conductances
were removed and replaced by matched spatial synaptic channels with the same
reversal potentials, activation parameters, time constants, and sampled
maximal conductances. Coherent population activity was quantified as
\[
    \mathrm{CP}(t)
    =
    \left\langle v_E(x,t) \right\rangle_x^2,
\]
and Figure~\ref{fig:PING_rhythm_comparison} displays
$\log_{10}[1+\mathrm{CP}(t)]$.

For Figure~\ref{fig:heterogeneity_tuninig}, the conductances
$g_l$, $\bar{g}_e$, and $\bar{g}_i$ were rescaled together according to
\begin{equation}
    (g_l,\bar{g}_e,\bar{g}_i)
    \longrightarrow
    \mu(g_l,\bar{g}_e,\bar{g}_i),
\end{equation}
with $\mu$ varied over $20$ equally spaced values between $0.1$ and $1.5$.
The rescaling was restricted to the central $10\%$, $20\%$, or $30\%$
of the spatial domain, while all other parameters were left unchanged.
For each condition, the excitatory population received the time-independent
input $I_{\mathrm{app},E}(x)=A_EG_8(x;0,0.5)$ with $A_E=0.30$, while the
inhibitory population received no direct input.

The population rhythm was quantified from the coherent spatial power
\begin{equation}
    \mathrm{CP}(t)
    =
    \left\langle
        \operatorname{clamp}\!\left(v_E(x,t),0,50\right)
    \right\rangle_x^2.
\end{equation}
The initial $30\%$ of each simulation was discarded as transient.
Oscillation peaks were detected using a minimum prominence equal to
$10\%$ of the range of $\mathrm{CP}(t)$ and a minimum separation of
$10$ sampled time points. A rhythm was retained only if at least three
complete inter-peak intervals were present and their coefficient of
variation did not exceed $0.5$. The population rhythm frequency was then
calculated as
\begin{equation}
    f_{\mathrm{pop}}
    =
    \frac{1000}
    {\left\langle\Delta t_{\mathrm{peak}}\right\rangle}
    \ \mathrm{Hz},
\end{equation}
with simulations failing these criteria assigned a frequency of zero.

\bibliographystyle{apalike}
{\small
\bibliography{CDC_2025}
}

\end{document}